\documentclass[aps,twocolumn,prb,preprintnumbers,amsmath,amssymb]{revtex4-1}
\usepackage{graphicx, bm}
\usepackage{dcolumn}
\usepackage{float}
\usepackage{latexsym}
\usepackage{amsmath}
\usepackage{graphics}
\usepackage{amssymb}
\usepackage{layout}
\usepackage{verbatim}
\usepackage{amsfonts,epsfig}
\usepackage{color}
\usepackage{setspace}
\newcommand{\beqnar}{\begin{eqnarray}}
\newcommand{\eeqnar}{\end{eqnarray}}

\newcommand{\beq}{\begin{equation}}
\newcommand{\eeq}{\end{equation}}

\begin{document}
\title{Graphene on SrTiO$_3$}
\author{S. Das Sarma and Qiuzi Li}
\affiliation{$^1$Condensed Matter Theory Center, Department of Physics, University of Maryland, College Park, Maryland 20742}
\date{\today}
\begin{abstract}
We study carrier transport through graphene on SrTiO$_3$ substrates by considering relative contributions of Coulomb and resonant impurity scattering to graphene resistivity.  We establish that charged impurity scattering must dominate graphene transport as the charge neutrality point is approached by lowering the carrier density, and in the higher density regime away from the neutrality point a dual model including both charged impurities and resonant defects gives an excellent description of graphene transport on SrTiO$_3$ substrates. We further establish that the non-universal high-density behavior of $\sigma(n)$ in different graphene samples on various substrates arises from the competition among different scattering mechanisms, and it is in principle entirely possible for graphene transport to be dominated by qualitatively different scattering mechanisms at high and low carrier densities.

\end{abstract}


\maketitle

\section{Introduction}

The substrate material has significant influence on graphene transport properties, which has been confirmed by many different experimental groups\cite{TanDasChen3,Jang_PRL08,Du_Sus08,BNDeanDas,Newaz}. The extensively used model \cite{dassarmaHwangAdam2,qzli_PRB11} of Coulomb disorder in the environment has earlier been found to provide a reasonable theoretical description for graphene transport on SiO$_2$ substrates from many different groups \cite{TanDasChen3} as well as on several other substrates \cite{AdamDasHwang4,Newaz,DasHwang_PRBBN11}. In a recent paper \cite{Couto_PRL11}, however, Couto {\it et al.} claim that the long-range Coulomb impurities do not play an important role based on their measured conductivity, $\sigma(n)$, of graphene at high carrier densities on  SrTiO$_3$ substrates. In addition, they argued that their transport data can be explained quantitatively by a so-called ``resonant scattering" model \cite{StauberFerreira5}, which gives the following expression for the carrier density $(n)$ dependence of the conductivity $\sigma$: $\sigma(n)=\frac{2 e^2 }{\pi h}\frac{n}{n_i}\ln^2(\sqrt{n \pi R^2})$, where $n_i, R$ are respectively the concentration and the range of the resonant scattering defects in graphene. Motivated by this experiment\cite{Couto_PRL11}, we revisit the question of the role of various types of disorder on different substrates in controlling the density-dependent conductivity of graphene, comparing, in particular, the low- and high-density conductivity limited by distinct scattering mechanisms.  In particular, the low-density graphene transport behavior on SrTiO$_3$ (or for that matter, on any other substrate) is likely to be always dominated by long-range Coulomb disorder, independent of the other scattering mechanisms (e.g. short-range disorder, resonant scattering) which may be operational at high carrier density. One important finding of our current work is to show that it is indeed possible for graphene transport to be determined by different scattering mechanisms at low and high carrier densities, a result which has been implicit in earlier works in this subject.

First, we note that a conductivity formula with $\sigma \sim \frac{n}{n_i}\ln^2(\sqrt{n/n_0})$ cannot, by definition, qualitatively account for the most important aspect of graphene transport, namely, the existence of the low-density minimum conductivity for a finite range of density around the Dirac (i.e. charge neutrality) point. Thus, the resonant scattering model, even in the most favorable circumstances, can only be a rather phenomenological data-fitting scheme for $\sigma(n)$ in an intermediate density range $n_c < n < n_0$ where $n_c$ defines the density regime for the graphene minimum conductivity plateau around the Dirac point (taken to be at $n=0$) and $n_0 \equiv (\pi R^2)^{-1}$. Second, as no physical evidence for the existence of these ``resonant scattering" short-range atomic defects (with $n_i \sim 3\times 10^{11}$ cm$^{-2}$) is presented in Ref. [\onlinecite{Couto_PRL11}] except for providing an intermediate-density phenomenological fit to their conductivity data, it is not clear whether the resonant scattering is the only scattering mechanism for graphene on SrTiO$_3$. 

One interesting observation in Ref.~[\onlinecite{Couto_PRL11}] is the apparent absence of much temperature dependence in the graphene conductivity while the substrate dielectric constant $\kappa (T)$ is changing substantially as a function of temperature as the substrate material, i.e. SrTiO$_3$ undergoes a paraelectric to ferroelectric transition with the lowering of temperature.  One possibility that cannot be ruled out in this context is that the substrate impurity density is also changing substantially during this temperature-induced substrate structural ferroelectric transition.  In the current work, we explicitly incorporate such a possibility in the theory for disorder limited graphene conductivity to compare with the data of Ref.~[\onlinecite{Couto_PRL11}]. We believe that the presence of some background Coulomb disorder is essential for understanding the low-density minimum conductivity behavior near the charge neutrality point in graphene on any substrate.

\begin{figure}
\includegraphics[width=0.99\columnwidth]{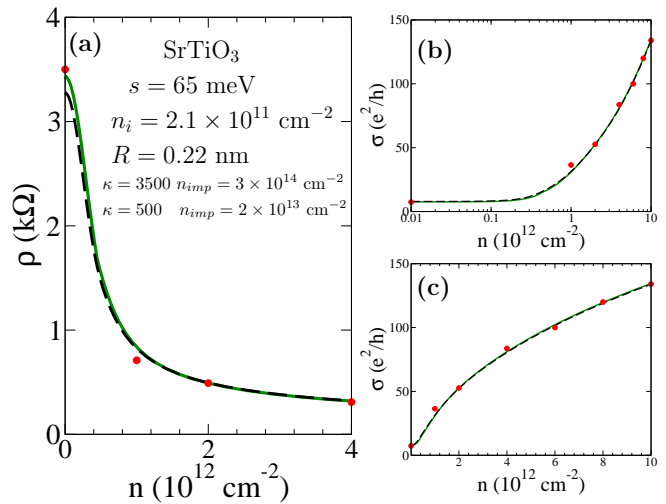}
\caption{Fits to the experimental data of graphene on SrTiO$_3$ (Fig. 2 of Ref.~[\onlinecite{Couto_PRL11}]) including Coulomb impurity $n_{imp}$ and resonant impurity $n_i$. (a) Resistance $\rho$ as a function of carrier density $n$. (b) Conductance $\sigma$ as a
function of $n$ (in semi-logarithmic scale). (c) Conductance $\sigma$ as a function of $n$ (in linear scale). The solid (dashed) line is for the temperature $T=0.25$ ($50$) K, $\kappa = 3500$ ($500$) and $n_{imp} = 3 \times 10^{14}$ ($2 \times 10^{13}$ cm$^{-2}$).  Note that we use the theory presented in Ref.~[\onlinecite{qzli_PRB11}], $s$ denotes the potential fluctuation associated with the puddles induced by Coulomb disorder \cite{YZhang_NP09}, the average distance of the charged impurity from the graphene sheet used in (a), (b) and (c) is $d=1$ \AA \ and the range of the resonant scattering defects $R = 0.22$ nm. Solid (red) dots represent the experimental data extracted from Ref.~[\onlinecite{Couto_PRL11}].
}
\label{fig:1}
\end{figure}

In this paper, we show our best theoretical fits to the data of Ref.~[\onlinecite{Couto_PRL11}] in Fig.~\ref{fig:1}, finding that a dual model involving both Coulomb and resonant scattering disorder can well explain the data of Ref.~[\onlinecite{Couto_PRL11}] with the single assumption of a variable background charged impurity density with varying temperature. The assumption of a temperature-dependent charged impurity density for SrTiO$_3$ substrates is not an arbitrary data fitting ploy because the complicated lattice ferroelectric properties of SrTiO$_3$ leading to the strong functional dependence of the dielectric constant on temperature may very well also produce a temperature dependent charged impurity density increasing strongly with decreasing temperature just as the actual carrier density in graphene on SrTiO$_3$ increases rapidly with decreasing temperature at a fixed gate voltage (see Fig. 2 in the Supplementary Information of Ref.~[\onlinecite{Couto_PRL11}]). 

In the remaining of this paper, we demonstrate that resonant scattering by itself can never explain the experimental data in the whole carrier density range for any substrate since the low-density minimum conductivity behavior cannot be understood based on a model which includes only resonant scattering. We present our best theoretical fits to the experimental data in Ref.~[\onlinecite{Couto_PRL11}] for SrTiO$_3$ using the new theory of long-range Coulomb disorder and the resonant scattering defects. We establish that a simple model of single type of scattering mechanism can not describe the transport data both at low and high carrier density for complex oxide substrates. We also compare three different models to explain transport properties of graphene on different substrates, finding that different behaviors of $\sigma(n)$ for different samples on different substrates are attributed to the competition among different scattering sources. The main message of our paper illustrated through concrete calculations, not emphasized explicitly in earlier works (although it might have been implicit), is that the high-density transport in graphene is nonuniversal and reflects the combination of various scattering processes arising from the substrate whereas the low-density transport is always dominated by long-range Coulomb disorder.  In addition to the SrTiO$_3$ substrates used in Ref.~[\onlinecite{Couto_PRL11}], we also consider graphene transport on several other common substrates (e.g. SiO$_2$, h-BN, and vacuum) to make our point about the universality (nonuniversality) of the transport behavior at low(high) densities.

\section{Theoretical formalism}

In this section, we first discuss the effects of various scattering mechanisms on the electronic transport of graphene. We will analyze four different sources of disorder covering the various realistic possibilities for resistive disorder scattering in graphene (a) randomly distributed long-range screened Coulomb impurity\cite{dassarmaHwangAdam2}, (b) short-range point defects\cite{dassarmaHwangAdam2}, (c) resonant disorder\cite{StauberFerreira5}, and (d) correlated long-range Coulomb impurity\cite{qzli_PRL11} in the following. The inverse of the total scattering time $1/\tau_{tot}$ is the sum of different inverse scattering times due to various scattering mechanisms at low temperatures. We further discuss three different models leading to the sub-linear behavior of $\sigma(n)$ of graphene on various substrates: (1) the standard model including randomly distributed Coulomb impurity and short-range scattering mechanism; (2) the resonant scattering mechanism; (3) the recently proposed correlated charged impurity model\cite{qzli_PRL11}. We note that at densities above the puddle-dominated density regime, the conductivity is strictly linear in carrier density for scattering by long-range Coulomb disorder.

First, the scattering time due to randomly distributed charged impurity, denoted as $\tau_{imp}$, is given by\cite{dassarmaHwangAdam2},
\begin{eqnarray}
\dfrac{\hbar}{\tau_{imp}(\epsilon_{p{\bf k}})}= 2\pi n_{imp} \int \frac{d^2 k'}{(2\pi)^2} \left|\frac{v_i(q)}{\varepsilon(q)}\right|^2 g(\theta_{\bf kk'})\nonumber \\
\times \left[1-\cos\theta_{\bf kk'}\right]\delta(\epsilon_{p\mathbf{k'}}-\epsilon_{p\mathbf{k}})
\label{eq:long}
\end{eqnarray}
where $\epsilon_{p\mathbf{k}} = p \hbar v_F k$ is the energy of non-interacting Dirac fermion with the Fermi velocity $v_F$ for the pseudospin state ``$p$" and 2D wave vector ${\bf k}$, $\theta_{\bf kk'}$ is
the scattering angle between in- and out- wave vectors
${\bf k}$ and $\bf k'$, $\mathbf{q}= \mathbf{k-k'}$, $g(\theta_{\bf kk'})=\left[1+\cos\theta_{\bf
    kk'}\right]/2$ is a wave function form factor associated with the
chiral matrix of monolayer graphene (and is determined by its band
dispersion relation). $n_{imp}$ is the 2D density of the randomly distributed screened Coulomb impurity\cite{DasEnrico_PRB10}. $v_i (q) =
2 \pi e^2/(\kappa q)$ is the Fourier transform of the two-dimensional (2D) Coulomb
potential in an effective background dielectric constant $\kappa$, $\varepsilon(q)$ is the static dielectric function of graphene within random-phase approximation (RPA)\cite{HwangDas_PRB07}. This long-range Coulomb disorder leads to the linear density dependent conductivity at low temperatures, \cite{dassarmaHwangAdam2}.
\begin{equation}
\sigma_{imp} = \frac{e^2}{h} \frac{n}{2 n_{imp} r_s^2 G_1 (r_s)}
\end{equation}
where $r_s = e^2/ (\hbar v_F \kappa)$ is graphene fine structure constant and $G_1 (x) = \frac{\pi}{4} + 6 x - 6 \pi x^2 + 4 x (6 x^2 -1) g (x)$ with $g(x) = \text{sech}^{-1}(2x)/ \sqrt{1-4 x^2}$ for $x< \frac{1}{2}$ and $\sec^{-1}(2x)/\sqrt{4 x^2 -1}$ for $x > \frac{1}{2}$.

We then provide the short-range disorder scattering time $\tau_{sd}$, given by\cite{dassarmaHwangAdam2}
\begin{eqnarray}
\dfrac{\hbar}{\tau_{sd}(\epsilon_{{\bf k}})}= \frac{k}{4 \hbar v_F} n_{sd} V_0^2
\label{eq:short}
\end{eqnarray}
where $n_{sd}$ is the 2D short-range impurity density and $V_0$ is a constant short-range
(i.e. a $\delta$-function in real space) potential strength. The conductivity at low temperature induced by this short-range impurity has the following form,
\begin{equation}
\sigma_{sd} = \frac{8 e^2}{h} \frac{(\hbar v_F)^2}{n_{sd} V_0^2}
\end{equation}
which is independent of carrier density ($\sigma(n) \sim $ constant).

The scattering time due to resonant defects $\tau_i$ has been shown to have the following form\cite{StauberFerreira5}
\begin{eqnarray}
\dfrac{1}{\tau_i(\epsilon_{{\bf k}})}= \frac{\pi^2 v_F n_i}{k(\ln kR)^2}
\label{eq:resonant}
\end{eqnarray}
where $n_i$ is the concentration of the resonant defects and $R$ is the potential range of the resonant scattering defects. The logarithmic term in the scattering time of resonant defects gives rise to the sublinear density dependent conductivity\cite{StauberFerreira5},
\begin{equation}
\sigma_i = \frac{2 e^2 }{\pi h}\frac{n}{n_i}\ln^2(\sqrt{n \pi R^2})
\end{equation}

We now turn to the discussion of the scattering time due to the spatially correlated charged impurity $\tau_c$ is given by\cite{qzli_PRL11}
\begin{eqnarray}
\dfrac{\hbar}{\tau_c(\epsilon_{p{\bf k}})} &=& 2\pi n_{imp} \int \dfrac{d^2 k'}{(2\pi)^2}\left|\frac{v_i(q)}{\varepsilon(q)}\right|^2 S(q) \nonumber \\
& \times & g(\theta_{\bf kk'}) \left[1-\cos\theta_{\bf kk'}\right]\delta(\epsilon_{p\mathbf{k'}}-\epsilon_{p\mathbf{k}})
\label{eq:mscatt}
\end{eqnarray}
where $S(q) =1-2\pi n_{imp}\dfrac{r_{0}}{q}J_{1}(qr_{0})$ is the structure factor using the simple continuum analytic model\cite{qzli_PRL11}, $J_1(x)$ is the Bessel function of the first kind and the correlation effects is defined by the length scale $r_0<r_i\equiv (\pi n_{imp})^{-1/2}$, the so-called correlation length. The asymptotic form of conductivity at low ``$k_F$'' arising from correlated long-range Coulomb disorder is found to be\cite{qzli_PRL11},
\begin{equation}
\sigma_c = \frac{e^2}{h} \frac{n}{2 n_{imp} r_s^2 G_1 (r_s)}\frac{1}{1- a + B a^2 n/n_{imp}}
\end{equation}
where $a = \pi n_{imp} r_0 ^2$ and $B = G_2(r_s)/[2 G_1(r_s)]$ with $G_2(y) = \frac{\pi}{16}- \frac{4 y}{3} + 3 \pi y^2 + 40 y^3[1- \pi y+ \frac{4}{5} (5 y^2 -1)]g(y)$. The correlated Coulomb disorder also leads to a sub-linear density-dependent conductivity, which was adopted to explain the enhancement of both sublinearity of $\sigma (n)$ and mobility reported in Ref.~[\onlinecite{fuhrer2010}].

Comparing different scattering mechanisms discussed above, it is obvious that the long-range Coulomb disorder dominates $\sigma (n)$ at low carrier density since the corresponding scattering rate is asymptotically the largest in the vanishing density limit, implying that as the density decreases, only Coulomb disorder would matter near the Dirac point. The long-range Coulomb disorder is also essential in order to explain the electron-hole puddle induced low-density minimum conductivity\cite{dassarmaHwangAdam2}. In addition to the resonant and correlated long-range Coulomb scattering, we note that a standard model involving both long-range ($\sigma(n) \sim n$) and short-range scatterers can explain the sublinearity of $\sigma(n)$ at higher carrier density, which has successfully explain the conductivity of graphene on boron nitride substrates\cite{DasHwang_PRBBN11}. We argue that the sample dependent $\sigma(n)$ at high carrier density may well arise from the competition between different scattering mechanisms.

\begin{figure}
\begin{center}
\includegraphics[width=0.99\columnwidth]{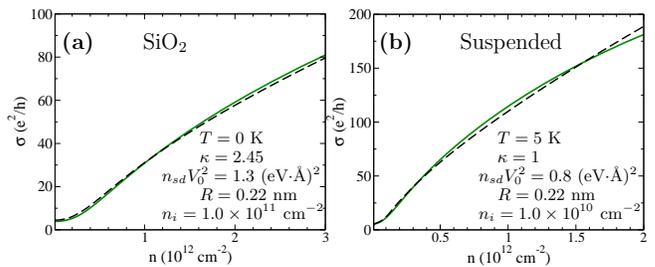}
  \caption{(Color online). Theoretical fits to $\sigma(n)$ for graphene on other substrates. The solid line is the fit including  Coulomb impurity $n_{imp}$ and short-range point defect $n_{sd} V_0^2$.  The dashed line is the fit including Coulomb impurity $n_{imp}$ and resonant impurity $n_i$. (a) Fits to the $\sigma(n)$ data for graphene on SiO$_2$ (Fig. 2 of Ref.~[\onlinecite{Jang_PRL08}]). The potential fluctuation $s=55$ meV, the effective dielectric constant $\kappa = 2.45$, and the average distance of the charged impurity from the graphene sheet is $d=1$ \AA. (b) Fits to the $\sigma(n)$ data of suspended graphene (Fig. 3(c) of Ref.~[\onlinecite{BolotinYacoby}]). The potential fluctuation $s=10$ meV, the temperature $T=5$ K, the dielectric constant $\kappa = 1.0$, and the average distance of the charged impurity from the graphene sheet is $d=1$ \AA. }
\label{fig:com2}
\end{center}
\end{figure}

\section{Numerical results}

In this section, we show our theoretical fits to some representative $\sigma(n)$ data for graphene on four different substrates. The analytical formula for $\sigma(n)$ shown in the last section are calculated for homogeneous systems. To capture the inhomogeneous nature of the graphene landscape close to the charge neutrality point (CNP), we assume a Gaussian form of potential fluctuation, parametrized by ``$s$'' and apply the effective medium theory of conductance with binary mixture of component, i.e., electron-hole puddles, the details of which have been given in Ref.~[\onlinecite{qzli_PRB11}]. The minimum conductivity plateau of the puddle dominated regime in graphene around the charge neutrality point is just well-captured in our theory.

In Fig.~\ref{fig:1}, we show our best theoretical fits to graphene transport data of Ref.~[\onlinecite{Couto_PRL11}], using a dual scattering models: the new theory of long-range Coulomb disorder and the resonant scattering defects, which give $\sigma(n)\sim n\ln^2(\sqrt{n \pi R^2})$.  The Coulomb disorder theory of transport follows Ref.~[\onlinecite{qzli_PRB11}] and includes the effect of Coulomb disorder-induced electron-hole puddles through the potential fluctuation parameter ``s". The new model explains the conductivity of graphene on SrTiO$_3$ in the whole range of carrier density. Couto {\it et al.} in Ref.~[\onlinecite{Couto_PRL11}] pointed out that at fixed value of external gate voltage, the carrier density $n$ and the dielectric constant of SrTiO$_3$ substrate would increase by approximately one order of magnitude as the temperature is lowered from $50$ K down to $250$ mK, which may also lead to the temperature-dependent charged impurity density. Resonant scattering may account for the conductivity of graphene on SrTiO$_3$ away from the Dirac point. Note, however, that Coulomb disorder induced electron-hole puddles, which give rise to the existence of minimum conductivity, dominate the transport properties of graphene on SrTiO$_3$ close to the Dirac point and thus some charged impurity scattering must be present in the samples of Ref.~[\onlinecite{Couto_PRL11}].

In Fig.~\ref{fig:com2}, we present the comparison of the $\sigma(n)$ results between the standard model and the new model for graphene on SiO$_2$ substrate \cite{Jang_PRL08} and the suspended graphene\cite{Bolotin_PRL08,BolotinYacoby}. Both models can fit the experimental data very well. Fig.~\ref{fig:com2} shows that the above mentioned two distinct dual scattering models  are equally successful in describing the experimental data for both SiO$_2$-based and suspended graphene, and there is no particular reason, within the transport data fitting scheme without additional information, to preferably choose one model over the other.  We mention that the parameters of the models ($n_{imp}$, $n_{sd} V_0^2$ in the standard model and $n_{imp}$, $n_i$ in the new model) are not absolutely unique, and it is possible to get equivalent fits by adjusting the parameter sets somewhat.  This is expected because $\sigma(n)$ is a smooth function at higher density and somewhat different parameter sets for disorder cannot be distinguished since the models have only qualitative and semi-quantitative predictive power.  We emphasize, however, that the resonant scattering model by itself cannot explain the minimum conductivity phenomenon around the charge neutrality point in graphene, existing all the way to the room temperature and above, which is a generic observation for graphene on all substrates.

\begin{figure}
\begin{center}
\includegraphics[width=0.99\columnwidth]{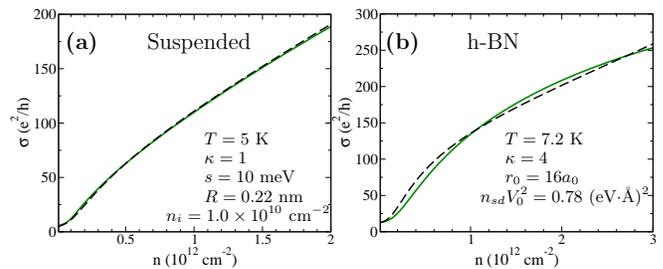}
  \caption{(Color online). (a) Fits to the $\sigma(n)$ data of suspended graphene (Fig. 3(c) of Ref.~[\onlinecite{BolotinYacoby}]). The solid line is the fit including  the Coulomb impurity $n_{imp}=0.15\times 10^{10}$ cm$^{-2}$ and resonant scattering  $n_{i}$.  The dashed line is the fit including correlated Coulomb impurity $n_{imp} = 1.6 \times 10^{10}$ cm$^{-2}$ and correlation length $r_0 = 60 a_0$ with $a_0=4.92$ \AA.  (b) Fits to the $\sigma(n)$ data for graphene on boron nitride (Fig. 3(a) of Ref.~[\onlinecite{BNDeanDas}]). The solid line is the fit including  Coulomb impurity $n_{imp}=1.4\times 10^{11}$ cm$^{-2}$ and short-range point defect $n_{sd} V_0^2$, the potential fluctuation $s=44$ meV.  The dashed line is the fit including correlated Coulomb impurity $n_{imp} = 3.7 \times 10^{11}$ cm$^{-2}$, and the potential fluctuation $s=38$ meV.  Note that the average distance of the charged impurity from the graphene sheet used in (a) and (b) is $d=0$ \AA.}
\label{fig:f3}
\end{center}
\end{figure}

In Fig.~\ref{fig:f3}(a), we compare the theoretical fits between the new model and the correlated disorder model proposed in Ref.~[\onlinecite{qzli_PRL11}] for suspended graphene.  Because the annealing procedure is routinely used in preparing suspended graphene samples, it is very likely to introduce some spatial correlation among the Coulomb impurities. In Fig.~\ref{fig:f3}(b), we compare the standard model with the correlated disorder model for graphene on hexagonal boron nitride (h-BN) substrate. Graphene on boron nitride may also induce some inter-impurity correlations imposed by the similarity between boron nitride  and graphene lattice structure. We know, from Fig.~\ref{fig:f3}, all of these three models can give reasonable fits to the experimental data of the density-dependent conductivity. They capture the experimentally observed strong-nonlinearity in the density-dependent conductivity $\sigma(n)$. The long-range Coulomb impurity leads to the density inhomogeneity in the graphene system, which is essential in explaining the conductivity  plateau close to the Dirac point.

\section{Discussion and conclusion}

Before concluding, we argue that the resonant scattering theory with $\sigma\sim\frac{n}{n_i}\ln^2(\sqrt{n/n_0})$ predicts a vanishing of graphene conductivity as the carrier density approaches the Dirac point ($n=0$) which is not observed in the experiments of Ref.~[\onlinecite{Couto_PRL11}], where $\sigma(n)$ becomes an approximate constant at low density around the Dirac point in complete contrast with the resonant scattering predictions and thus, the resonant scattering theory cannot obviously be the complete story underlying the transport mechanisms in Ref.~[\onlinecite{Couto_PRL11}].  We emphasize that the high-density behavior of $\sigma(n)$ in graphene is, in general, known \cite{dassarmaHwangAdam2} to be nonuniversal with different samples on different substrates showing different behaviors \cite{TanDasChen3} arising from the competition among Coulomb disorder, short-range disorder, ripples, and (perhaps even) resonant scattering disorder. It is entirely possible that the transport data of Ref.~[\onlinecite{Couto_PRL11}] is best described by a combination of Coulomb disorder and resonant scattering, where the low-density minimum conductivity arises from the Coulomb disorder and the intermediate (sublinear in) density conductivity arises from resonant scattering as shown in Fig.\ref{fig:1}, assuming dual independent scattering by charged impurity and resonant scattering centers. The fact that we get excellent agreement over the whole density range of the experimental data indicates that our dual scattering model is a more reasonable description than the pure resonant scattering model. We have also obtained similar good theoretical fits to the existing graphene data as shown in Fig.~\ref{fig:com2} and \ref{fig:f3}. Thus, the possibility that the resonant scattering mechanism is operational at some level in graphene transport at higher densities on any substrate (not just SrTiO$_3$) cannot definitively be ruled out as we show in this paper by considering several different substrates.  What can be stated rather definitively is that the low-density transport must always be dominated in graphene by charged impurity scattering.

We conclude by emphasizing that the resonant scattering model by itself can not explain the transport data in graphene. In particular, the long-range charged impurity scattering is needed in order to explain the existence of the minimum conductivity close to the charge neutrality point. The main point of our paper is that the understanding of the graphene minimum conductivity phenomenon necessarily requires the inclusion of scattering by random charged impurities in the environment. We emphasize that since $\sigma_r/\sigma_c \sim \ln^2(\sqrt{n/n_0})$, where $\sigma_{r,c}$ are the resonant scattering and Coulomb scattering induced conductivity respectively, Coulomb disorder\cite{dassarmaHwangAdam2}, with $\sigma_c \sim n$ , must necessarily dominate graphene resistivity as one approaches the $n \rightarrow 0$ charge neutrality point, not just in SrTiO$_3$, but in all systems. The high density graphene transport is in general, however, nonuniversal where many scattering mechanisms may contribute depending on the density range.

\begin{acknowledgments}

We thank Drs. Shaffique Adam, Euyheon Hwang, and Enrico Rossi for discussions.  This work is supported by ONR-MURI.

\end{acknowledgments}


\begin{thebibliography}{10}%
\makeatletter
\providecommand \@ifxundefined [1]{%
 \ifx #1\undefined \expandafter \@firstoftwo
 \else \expandafter \@secondoftwo
\fi
}%
\providecommand \@ifnum [1]{%
 \ifnum #1\expandafter \@firstoftwo
 \else \expandafter \@secondoftwo
\fi
}%
\providecommand \enquote [1]{``#1''}%
\providecommand \bibnamefont  [1]{#1}%
\providecommand \bibfnamefont [1]{#1}%
\providecommand \citenamefont [1]{#1}%
\providecommand\href[0]{\@sanitize\@href}%
\providecommand\@href[1]{\endgroup\@@startlink{#1}\endgroup\@@href}%
\providecommand\@@href[1]{#1\@@endlink}%
\providecommand \@sanitize [0]{\begingroup\catcode`\&12\catcode`\#12\relax}%
\@ifxundefined \pdfoutput {\@firstoftwo}{%
 \@ifnum{\z@=\pdfoutput}{\@firstoftwo}{\@secondoftwo}%
}{%
 \providecommand\@@startlink[1]{\leavevmode}%
 \providecommand\@@endlink[0]{}%
}{%
 \providecommand\@@startlink[1]{%
  \leavevmode
  \pdfstartlink
   attr{/Border[0 0 1 ]/H/I/C[0 1 1]}%
   user{/Subtype/Link/A<</Type/Action/S/URI/URI(#1)>>}%
  \relax
 }%
 \providecommand\@@endlink[0]{\pdfendlink}%
}%
\providecommand \url  [0]{\begingroup\@sanitize \@url }%
\providecommand \@url [1]{\endgroup\@href {#1}{\urlprefix}}%
\providecommand \urlprefix [0]{URL }%
\providecommand \Eprint[0]{\href }%
\@ifxundefined \urlstyle {%
  \providecommand \doi [1]{doi:\discretionary{}{}{}#1}%
}{%
  \providecommand \doi [0]{doi:\discretionary{}{}{}\begingroup
  \urlstyle{rm}\Url }%
}%
\providecommand \doibase [0]{http://dx.doi.org/}%
\providecommand \Doi[1]{\href{\doibase#1}}%
\providecommand \bibAnnote [3]{%
  \BibitemShut{#1}%
  \begin{quotation}\noindent
    \textsc{Key:}\ #2\\\textsc{Annotation:}\ #3%
  \end{quotation}%
}%
\providecommand \bibAnnoteFile [2]{%
  \IfFileExists{#2}{\bibAnnote {#1} {#2} {\input{#2}}}{}%
}%
\providecommand \typeout [0]{\immediate \write \m@ne }%
\providecommand \selectlanguage [0]{\@gobble}%
\providecommand \bibinfo [0]{\@secondoftwo}%
\providecommand \bibfield [0]{\@secondoftwo}%
\providecommand \translation [1]{[#1]}%
\providecommand \BibitemOpen[0]{}%
\providecommand \bibitemStop [0]{}%
\providecommand \bibitemNoStop [0]{.\EOS\space}%
\providecommand \EOS [0]{\spacefactor3000\relax}%
\providecommand \BibitemShut [1]{\csname bibitem#1\endcsname}%
\bibitem{TanDasChen3}%
  \BibitemOpen
  \bibfield{author}{%
  \bibinfo {author} {\bibfnamefont{Y.-W.}\ \bibnamefont{Tan~{\it et al.}}},\ }%
  \bibfield{journal}{%
  \bibinfo {journal} {Phys. Rev. Lett.}\ }%
  \textbf{\bibinfo {volume} {99}},\ \bibinfo {pages} {246803 (2007); J. H. Chen
  {\it et al.}, Nat. Phys. {\bf 4}, 377 (2008); F. Chen,{\it et al.}, Nano
  Lett. {\bf 9}, 1621 (2009); W. Zhu,{\it et al.}, Phys. Rev. B {\bf 80},
  235402 (2009); X. Hong,{\it et al.}, Phys. Rev. B {\bf 80}, 241415 (2009); K.
  Zou,{\it et al.}, Phys. Rev. B {\bf 82}, 081407 (2010); J. Heo,{\it et al.},
  Phys. Rev. B {\bf 84}, 035421 (2011); D. Farmer,{\it et al.}, Phys. Rev. B
  {\bf 84}, 205417} (\bibinfo {year} {2011})%
  \bibAnnoteFile{NoStop}{TanDasChen3}%
\bibitem{Jang_PRL08}%
  \BibitemOpen
  \bibfield{author}{%
  \bibinfo {author} {\bibfnamefont{C.}~\bibnamefont{Jang}}, \bibinfo {author}
  {\bibfnamefont{S.}~\bibnamefont{Adam}}, \bibinfo {author}
  {\bibfnamefont{J.-H.}\ \bibnamefont{Chen}}, \bibinfo {author}
  {\bibfnamefont{E.~D.}\ \bibnamefont{Williams}}, \bibinfo {author}
  {\bibfnamefont{S.}~\bibnamefont{Das~Sarma}},\ and\ \bibinfo {author}
  {\bibfnamefont{M.~S.}\ \bibnamefont{Fuhrer}},\ }%
  \bibfield{journal}{%
  \bibinfo {journal} {Phys. Rev. Lett.}\ }%
  \textbf{\bibinfo {volume} {101}},\ \bibinfo {pages} {146805} (\bibinfo {year}
  {2008})%
  \bibAnnoteFile{NoStop}{Jang_PRL08}%
\bibitem{Du_Sus08}%
  \BibitemOpen
  \bibfield{author}{%
  \bibinfo {author} {\bibfnamefont{X.}~\bibnamefont{Du}}, \bibinfo {author}
  {\bibfnamefont{I.}~\bibnamefont{Skachko}}, \bibinfo {author}
  {\bibfnamefont{A.}~\bibnamefont{Barker}},\ and\ \bibinfo {author}
  {\bibfnamefont{E.~Y.}\ \bibnamefont{Andrei}},\ }%
  \bibfield{journal}{%
  \bibinfo {journal} {Nat. Nanotech.}\ }%
  \textbf{\bibinfo {volume} {3}},\ \bibinfo {pages} {491} (\bibinfo {year}
  {2008})%
  \bibAnnoteFile{NoStop}{Du_Sus08}%
\bibitem{BNDeanDas}%
  \BibitemOpen
  \bibfield{author}{%
  \bibinfo {author} {\bibfnamefont{C.~R.}\ \bibnamefont{Dean}}, \bibinfo
  {author} {\bibfnamefont{A.~F.}\ \bibnamefont{Young}}, \bibinfo {author}
  {\bibfnamefont{I.}~\bibnamefont{Meric}}, \bibinfo {author}
  {\bibfnamefont{C.}~\bibnamefont{Lee}}, \bibinfo {author}
  {\bibfnamefont{L.}~\bibnamefont{Wang}}, \bibinfo {author}
  {\bibfnamefont{S.}~\bibnamefont{Sorgenfrei}}, \bibinfo {author}
  {\bibfnamefont{K.}~\bibnamefont{Watanabe}}, \bibinfo {author}
  {\bibfnamefont{T.}~\bibnamefont{Taniguchi}}, \bibinfo {author}
  {\bibfnamefont{P.}~\bibnamefont{Kim}}, \bibinfo {author}
  {\bibfnamefont{K.~L.}\ \bibnamefont{Shepard}},\ and\ \bibinfo {author}
  {\bibfnamefont{J.}~\bibnamefont{Hone}},\ }%
  \bibfield{journal}{%
  \bibinfo {journal} {Nat. Nanotechnol.}\ }%
  \textbf{\bibinfo {volume} {5}},\ \bibinfo {pages} {722} (\bibinfo {year}
  {2010})%
  \bibAnnoteFile{NoStop}{BNDeanDas}%
\bibitem{Newaz}%
  \BibitemOpen
  \bibfield{author}{%
  \bibinfo {author} {\bibfnamefont{A.~K.~M.}\ \bibnamefont{Newaz}}, \bibinfo
  {author} {\bibfnamefont{Y.~S.}\ \bibnamefont{Puzyrev}}, \bibinfo {author}
  {\bibfnamefont{B.}~\bibnamefont{Wang}}, \bibinfo {author}
  {\bibfnamefont{S.~T.}\ \bibnamefont{Pantelides}},\ and\ \bibinfo {author}
  {\bibfnamefont{K.~I.}\ \bibnamefont{Bolotin}},\ }%
  \bibfield{journal}{%
  \bibinfo {journal} {Nat. Commun.}\ }%
  \textbf{\bibinfo {volume} {3}},\ \bibinfo {pages} {734} (\bibinfo {year}
  {2012})%
  \bibAnnoteFile{NoStop}{Newaz}%
\bibitem{dassarmaHwangAdam2}%
  \BibitemOpen
  \bibfield{author}{%
  \bibinfo {author} {\bibfnamefont{S.}~\bibnamefont{Das Sarma~{\it et al.}}},\
  }%
  \bibfield{journal}{%
  \bibinfo {journal} {Rev. Mod. Phys.}\ }%
  \textbf{\bibinfo {volume} {83}},\ \bibinfo {pages} {407 (2011); E. H.
  Hwang,{\it et al.}, Phys. Rev. Lett. {\bf 98}, 186806 (2007); S. Adam,{\it et
  al.}, Proc.\ Natl.\ Acad.\ Sci.\ USA {\bf 104}, 18392 (2007); E. Rossi,{\it
  et al.}, Phys. Rev. B {\bf 79}, 245423} (\bibinfo {year} {2009})%
  \bibAnnoteFile{NoStop}{dassarmaHwangAdam2}%
\bibitem{qzli_PRB11}%
  \BibitemOpen
  \bibfield{author}{%
  \bibinfo {author} {\bibfnamefont{Q.}~\bibnamefont{Li}}, \bibinfo {author}
  {\bibfnamefont{E.~H.}\ \bibnamefont{Hwang}},\ and\ \bibinfo {author}
  {\bibfnamefont{S.}~\bibnamefont{Das~Sarma}},\ }%
  \bibfield{journal}{%
  \bibinfo {journal} {Phys. Rev. B}\ }%
  \textbf{\bibinfo {volume} {84}},\ \bibinfo {pages} {115442} (\bibinfo {year}
  {2011})%
  \bibAnnoteFile{NoStop}{qzli_PRB11}%
\bibitem{AdamDasHwang4}%
  \BibitemOpen
  \bibfield{author}{%
  \bibinfo {author} {\bibfnamefont{S.}~\bibnamefont{Adam}}\ and\ \bibinfo
  {author} {\bibfnamefont{S.}~\bibnamefont{Das~Sarma}},\ }%
  \bibfield{journal}{%
  \bibinfo {journal} {Solid State Commun.}\ }%
  \textbf{\bibinfo {volume} {146}},\ \bibinfo {pages} {356 (2008); A. K. M.
  Newaz, {\it et al.}, Nat. Commun. {\bf 3}, 734} (\bibinfo {year} {2012})%
  \bibAnnoteFile{NoStop}{AdamDasHwang4}%
\bibitem{DasHwang_PRBBN11}%
  \BibitemOpen
  \bibfield{author}{%
  \bibinfo {author} {\bibfnamefont{S.}~\bibnamefont{Das~Sarma}}\ and\ \bibinfo
  {author} {\bibfnamefont{E.~H.}\ \bibnamefont{Hwang}},\ }%
  \bibfield{journal}{%
  \bibinfo {journal} {Phys. Rev. B}\ }%
  \textbf{\bibinfo {volume} {83}},\ \bibinfo {pages} {121405} (\bibinfo {year}
  {2011})%
  \bibAnnoteFile{NoStop}{DasHwang_PRBBN11}%
\bibitem{Couto_PRL11}%
  \BibitemOpen
  \bibfield{author}{%
  \bibinfo {author} {\bibfnamefont{N.~J.~G.}\ \bibnamefont{Couto}}, \bibinfo
  {author} {\bibfnamefont{B.}~\bibnamefont{Sac\'ep\'e}},\ and\ \bibinfo
  {author} {\bibfnamefont{A.~F.}\ \bibnamefont{Morpurgo}},\ }%
  \bibfield{journal}{%
  \bibinfo {journal} {Phys. Rev. Lett.}\ }%
  \textbf{\bibinfo {volume} {107}},\ \bibinfo {pages} {225501} (\bibinfo {year}
  {2011})%
  \bibAnnoteFile{NoStop}{Couto_PRL11}%
\bibitem{StauberFerreira5}%
  \BibitemOpen
  \bibfield{author}{%
  \bibinfo {author} {\bibfnamefont{T.}~\bibnamefont{Stauber~{\it et al.}}},\ }%
  \bibfield{journal}{%
  \bibinfo {journal} {Phys. Rev. B}\ }%
  \textbf{\bibinfo {volume} {76}},\ \bibinfo {pages} {205423 (2007); A.
  Ferreira,{\it et al.}, Phys. Rev. B {\bf 83}, 165402} (\bibinfo {year}
  {2011})%
  \bibAnnoteFile{NoStop}{StauberFerreira5}%
\bibitem{YZhang_NP09}%
  \BibitemOpen
  \bibfield{author}{%
  \bibinfo {author} {\bibfnamefont{J.}~\bibnamefont{Martin~{\it et al.}}},\ }%
  \bibfield{journal}{%
  \bibinfo {journal} {Nat. Phys.}\ }%
  \textbf{\bibinfo {volume} {4}},\ \bibinfo {pages} {144 (2008); E. Rossi,{\it
  et al.}, Phys. Rev. Lett. {\bf 101}, 166803 (2008); Y. Zhang,{\it et al.},
  Nat. Phys. {\bf 5}, 722 (2009); A. Deshpande,{\it et al.}, Phys. Rev. B {\bf
  83}, 155409} (\bibinfo {year} {2011})%
  \bibAnnoteFile{NoStop}{YZhang_NP09}%
\bibitem{qzli_PRL11}%
  \BibitemOpen
  \bibfield{author}{%
  \bibinfo {author} {\bibfnamefont{Q.}~\bibnamefont{Li}}, \bibinfo {author}
  {\bibfnamefont{E.~H.}\ \bibnamefont{Hwang}}, \bibinfo {author}
  {\bibfnamefont{E.}~\bibnamefont{Rossi}},\ and\ \bibinfo {author}
  {\bibfnamefont{S.}~\bibnamefont{Das~Sarma}},\ }%
  \bibfield{journal}{%
  \bibinfo {journal} {Phys. Rev. Lett.}\ }%
  \textbf{\bibinfo {volume} {107}},\ \bibinfo {pages} {156601 (2011); Q. Li, E.
  H. Hwang, and E. Rossi, , Solid State Commun. {\bf 152}, 1390} (\bibinfo
  {year} {2012})%
  \bibAnnoteFile{NoStop}{qzli_PRL11}%
\bibitem{DasEnrico_PRB10}%
  \BibitemOpen
  \bibfield{author}{%
  \bibinfo {author} {\bibfnamefont{S.}~\bibnamefont{Das~Sarma}}, \bibinfo
  {author} {\bibfnamefont{E.~H.}\ \bibnamefont{Hwang}},\ and\ \bibinfo {author}
  {\bibfnamefont{E.}~\bibnamefont{Rossi}},\ }%
  \bibfield{journal}{%
  \bibinfo {journal} {Phys. Rev. B}\ }%
  \textbf{\bibinfo {volume} {81}},\ \bibinfo {pages} {161407} (\bibinfo {year}
  {2010})%
  \bibAnnoteFile{NoStop}{DasEnrico_PRB10}%
\bibitem{HwangDas_PRB07}%
  \BibitemOpen
  \bibfield{author}{%
  \bibinfo {author} {\bibfnamefont{E.~H.}\ \bibnamefont{Hwang}}\ and\ \bibinfo
  {author} {\bibfnamefont{S.}~\bibnamefont{Das~Sarma}},\ }%
  \bibfield{journal}{%
  \bibinfo {journal} {Phys. Rev. B}\ }%
  \textbf{\bibinfo {volume} {75}},\ \bibinfo {pages} {205418} (\bibinfo {year}
  {2007})%
  \bibAnnoteFile{NoStop}{HwangDas_PRB07}%
\bibitem{fuhrer2010}%
  \BibitemOpen
  \bibfield{author}{%
  \bibinfo {author} {\bibfnamefont{J.}~\bibnamefont{Yan}}\ and\ \bibinfo
  {author} {\bibfnamefont{M.~S.}\ \bibnamefont{Fuhrer}},\ }%
  \bibfield{journal}{%
  \bibinfo {journal} {Phys. Rev. Lett.}\ }%
  \textbf{\bibinfo {volume} {107}},\ \bibinfo {pages} {206601} (\bibinfo {year}
  {2011})%
  \bibAnnoteFile{NoStop}{fuhrer2010}%
\bibitem{BolotinYacoby}%
  \BibitemOpen
  \bibfield{author}{%
  \bibinfo {author} {\bibfnamefont{K.}~\bibnamefont{Bolotin}}, \bibinfo
  {author} {\bibfnamefont{K.}~\bibnamefont{Sikes}}, \bibinfo {author}
  {\bibfnamefont{Z.}~\bibnamefont{Jiang}}, \bibinfo {author}
  {\bibfnamefont{M.}~\bibnamefont{Klima}}, \bibinfo {author}
  {\bibfnamefont{G.}~\bibnamefont{Fudenberg}}, \bibinfo {author}
  {\bibfnamefont{J.}~\bibnamefont{Hone}}, \bibinfo {author}
  {\bibfnamefont{P.}~\bibnamefont{Kim}},\ and\ \bibinfo {author}
  {\bibfnamefont{H.}~\bibnamefont{Stormer}},\ }%
  \bibfield{journal}{%
  \bibinfo {journal} {Solid State Commun.}\ }%
  \textbf{\bibinfo {volume} {146}},\ \bibinfo {pages} {351} (\bibinfo {year}
  {2008})%
  \bibAnnoteFile{NoStop}{BolotinYacoby}%
\bibitem{Bolotin_PRL08}%
  \BibitemOpen
  \bibfield{author}{%
  \bibinfo {author} {\bibfnamefont{K.~I.}\ \bibnamefont{Bolotin}}, \bibinfo
  {author} {\bibfnamefont{K.~J.}\ \bibnamefont{Sikes}}, \bibinfo {author}
  {\bibfnamefont{J.}~\bibnamefont{Hone}}, \bibinfo {author}
  {\bibfnamefont{H.~L.}\ \bibnamefont{Stormer}},\ and\ \bibinfo {author}
  {\bibfnamefont{P.}~\bibnamefont{Kim}},\ }%
  \bibfield{journal}{%
  \bibinfo {journal} {Phys. Rev. Lett.}\ }%
  \textbf{\bibinfo {volume} {101}},\ \bibinfo {pages} {096802} (\bibinfo {year}
  {2008})%
  \bibAnnoteFile{NoStop}{Bolotin_PRL08}%
\end{thebibliography}
\end{document}